\documentclass[10pt]{article}
\usepackage{amssymb}
\usepackage{mathrsfs}
\usepackage{multicol}
\usepackage{amsmath}
\usepackage{graphicx}
\usepackage{float}
\usepackage{flafter}
\usepackage{amsmath}
\usepackage{latexsym}
\usepackage{cite}
\usepackage{bm}
\usepackage{indentfirst}
\usepackage{balance}

\newcommand{\bee}{\begin{equation}}
\newcommand{\ee}{\end{equation}}
\newcommand{\beea}{\begin{eqnarray}}
\newcommand{\eea}{\end{eqnarray}}

\begin{document}

\title{Quantum magnetotransport properties of topological insulators under
strain}
\author{Ning Ma\thanks{%
Author to whom correspondence should be addressed. Electronic mail:
maning@tyut.edu.cn} \\
{\small Department of Physics, MOE Key Laboratory of Advanced Transducers and%
} \\
{\small Intelligent Control System, Taiyuan University of Technology,
Taiyuan 030024, China}}
\date{ }
\maketitle

\begin{abstract}
Recent experiments reveal that the strained bulk HgTe can be regard as a
three-dimensional topological insulator (TI). Motivated by this, we explore
the strain effects on the magnetotransport properties of the HgTe surface
states at magnetic field. We analytically derive the zero frequency Hall and
collisional conductivities, and find that the substrate induced strain
associated with the surface index of carriers, can result in the well
seperated surface quantum Hall plateaus and Shubnikov-de Haas oscillations.
These effects can be used to generate and detect surface polarization. 
\end{abstract}



\section{Introduction}

Recently, two- (2D) and three-dimensional (3D) topological insulators (TIs)
have drawn much attention in condensed matter physics \cite%
{1,2,3,4,5,6,7,8,9,10,11}. The relevant researches on 3D TIs mostly focus on
Bi$_{2}$Te$_{3}$, Bi$_{2}$Se$_{3}$, and Sb$_{2}$Te$_{3}$ compounds, which
possess a bulk energy gap and gapless conducting surface states \cite{7,8,9}%
. Thereinto, the surface states arise from the mismatch of the bulk
topological invariants on two sides of the surface, and can be regarded as
the 2D nonideal Dirac fermions with a single Dirac cone at high symmetry
points in the first Brillouin zone \cite{7,8,10}. These unique properties
account for the common features with graphene \cite{1,3,12}. However, such
3D TIs have strong defect doping and low carrier mobility in experiment, so
that the bulk conductivity always obscures the surface charge transport.
Typically, the predicted quantized magnetoelectric effect \cite{13,14} and
the surface Majorana fermions \cite{15}, can be found only when bulk
carriers are negligible compared to the surface states. Hence,
experimentally reaching the intrinsic TI regime, where bulk carriers are
absent, is now the central focus of the field.

While so far the focus has been on the above mentioned compounds, recently
growing attention is being paid to HgTe quantum wells \cite{16}, in which
the TI surface states were firstly predicted and observed \cite{1,2}. Bulk
HgTe also has Dirac-like surface states (Refs.~17 and 18) that originate
from the inversion between $\Gamma _{6}$ electron and $\Gamma _{8}$
light-hole bands, while the bulk band of $\Gamma _{8}$ heavy-hole coexists
with the surface state band, so that the surface states are always coupled
by metallic bulk states \cite{17,18}. This means that 3D HgTe is a semimetal
and thus not a TI in the strict sense. With strain applied, a bulk
insulating gap ($\sim 22$ meV) opens up at the touching point between the
light- and heavy-hole $\Gamma _{8}$ bands, and accordingly the strained bulk
HgTe becomes the real TI \cite{3,18}. Further transport measurements for the
strained HgTe on CdTe substrate exhibit the Rashba-like splitting induced by
the inversion symmetry breaking in a magnetic field \cite{18,19}. Also, the
Landau levels (LLs) are found to remain degenerate as long as the
hybridization can be neglected between the top and bottom surface states
(e.g. 70-nm-thick HgTe).

Motivated by this, we have theoretically investigate the strain effects on
the Shubnikov--de Haas (SdH) oscillations and Hall plateaus in the zero
frequency (dc) collisional and Hall conductivities using the Kubo formalism.
These magnetic oscillations appear due to the interplay of the quantum LLs
with the Fermi energy, and serve as a powerful technique to investigate the
Fermi surface and the spectrum of electron excitations. Our findings show
that the substrate induced strain could remove the LLs' surface degeneracy
in inversion symmetric Dirac cones on the top and bottom surfaces, which is
supported by HgTe transport experiments \cite{18,19}. Thus, the Dirac
particles of different surfaces present the well seperated quantum Hall and
SdH effects with different amplitudes and phases. Accordingly, this gives
rise to the splitting of LLs and the asymmetric spectrum of the
conductivity, in company with the mixture of LLs. Furthermore, we clarify
the connections of the SdH and Hall conductivities for different surfaces to
the abnormal integer Hall plateaus and SdH beating pattern, and the strain
induced changes in the zero-mode conductivity at different surfaces, etc.
These phenomena, absent in a conventional 2D electron gas (2DEG) and even in
graphene \cite{20,21,22,23,24,25,26}, should be attributed to the anomalous
spectrum of surface states in a fully stained TI. Our results are general
and can also be applied to Bi$_{2}$Se$_{3}$, Sb$_{2}$Te$_{3}$, and Bi$_{2}$Te%
$_{3}$, which would be very great news and certainly meet much interest by
the experimental groups.

\section{Results and discussion}

The surface states moves in the x-y plane under strain $\left( \Delta
\right) $ induced by the substrate, subjected to an external magnetic field $%
\mathbf{B=}\left( 0,0,B\right) ]$. Here we use the two Dirac cones model to
describe the experiments, and the 2D nonideal Dirac quasiparticle
Hamiltonian reads%
\begin{equation}
H=\mathbf{\tau }_{z}\upsilon _{F}\left( \mathbf{\sigma }_{x}\mathbf{\pi }%
_{y}-\mathbf{\sigma }_{y}\mathbf{\pi }_{x}\right) +\mathbf{\tau }_{z}\mathbf{%
I}\Delta .
\end{equation}%
In Eq.~$\left( 1\right) $, the first term arises from the spin--orbit
coupling (SOC). Due to the strong SOC, the TIs exhibit a unique
spin-momentum locking, which is essential for modeling topologically
nontrivial insulators. The second term results from the strain energy with $%
\mathbf{I}$ for the identity matrix and $\mathbf{\tau }_{z}=\pm 1$ for the
two surfaces of top {(facing vacuum)} and bottom (at CdTe interface). The
Fermi velocity $\upsilon _{F}$ of surface states in mercury telluride is $%
4.0\times 10^{5}$ m/s smaller than that in graphene ($\upsilon
_{F}=1.0\times 10^{6}$ m/s), $\left\{ \mathbf{\sigma }_{x},\mathbf{\sigma }%
_{y}\right\} $ is the vector of spin Pauli matrices, and $\mathbf{\pi =p+}e%
\mathbf{A/}c$ is the canonical momentum with $c$ for the speed of light and $%
\mathbf{A}$ for the vector potential yielding Landau gauge $\mathbf{A}%
=\left( 0,Bx,0\right) $. The resulting eigenvalues are \cite{18,19}%
\begin{equation}
E_{n,\lambda }^{\mathbf{\tau }_{z}}=\{%
\begin{array}{c}
\lambda \hbar \omega _{c}\sqrt{n}+\mathbf{\tau }_{z}\Delta ,n>0, \\
\mathbf{\tau }_{z}\Delta ,\text{\ \ \ \ \ \ \ \ \ \ \ }n=0.%
\end{array}%
\end{equation}%
with $\lambda $ for the electron ($\lambda =+1$) and hole ($\lambda =-1$)
bands. Integer $n$ $(n=0,1,2,...)$ represents the LLs $(n=0,1,2,...)$. The
cyclotron frequency is given by $\omega _{c}=\sqrt{2}\upsilon _{F}/\ell _{c}$
with $\ell _{c}=\sqrt{\hbar /eB}$ for the magnetic length. The
eigenfunctions of $\Psi _{n,\lambda }^{\tau _{z}}\left( r\right) $ reads
\begin{eqnarray}
\Psi _{n,+1}^{\tau _{z}}\left( r\right) &=&\frac{\exp \left( ik_{y}y\right)
}{\sqrt{L_{y}}}\left(
\begin{array}{c}
\alpha _{n}\Phi _{n-1}\left( \xi \right) \\
-\beta _{n}\Phi _{n}\left( \xi \right)%
\end{array}%
\right) ,  \notag \\
\Psi _{n,-1}^{\tau _{z}}\left( r\right) &=&\frac{\exp \left( ik_{y}y\right)
}{\sqrt{L_{y}}}\left(
\begin{array}{c}
\beta _{n}\Phi _{n-1}\left( \xi \right) \\
\alpha _{n}\Phi _{n}\left( \xi \right)%
\end{array}%
\right) ,
\end{eqnarray}%
in which $k_{y}=2\pi l/L_{y}$ ($l=0,1,2,...$) is the quantum number
corresponding to the translation symmetry along the $y$ axis with $L_{y}$
for the size of the surface in $y$ direction. The prefactors $\alpha _{n}$
and $\beta _{n}$ is, respectively, the cosine and sine of $\Theta /2$ with $%
\Theta =\arctan \left[ \omega _{c}\sqrt{n/2}/\left( \tau _{z}\Delta \right) %
\right] $ as%
\begin{eqnarray}
\alpha _{n} &=&\sqrt{\frac{F_{\lambda n}+\tau _{z}\Delta }{2F_{\lambda n}}},
\notag \\
\beta _{n} &=&\sqrt{\frac{F_{\lambda n}-\tau _{z}\Delta }{2F_{\lambda n}},}
\end{eqnarray}%
where $F_{\lambda n}=\lambda F_{n}$ with $F_{n}=\sqrt{n\hbar ^{2}\omega
_{c}^{2}+\Delta ^{2}}$. The harmonic oscillator eigenfunctions $\Phi _{n}$
are expressed in the normalized Hermitian polynomials $H_{n}\left( \xi
\right) $ as $\Phi _{n}=H_{n}\left( \xi \right) \exp \left( -\xi
^{2}/2\right) /\sqrt{2^{n}n!\ell _{c}\sqrt{\pi }}$, where we have $\xi
=\left( x+x_{c}\right) /\ell _{c}$ with $x_{c}=-\ell _{c}^{2}k_{y}$ for the
location of TI states in $\mathbf{x}$.

Assuming the electrons are elastically scattered by randomly distributed
charged impurities, we calculate the dc collisional conductivity following
the approaches \cite{27,28} as%
\begin{eqnarray}
\sigma _{xx} &=&\frac{\beta e^{2}}{S}\sum\limits_{\zeta ,\zeta ^{\prime
}}f\left( E_{\zeta }\right) \left[ 1-f\left( E_{\zeta ^{\prime }}\right) %
\right]  \notag \\
&&\times W_{\zeta \zeta ^{\prime }}\left( E_{\zeta },E_{\zeta ^{\prime
}}\right) \left( x_{\zeta }-x_{\zeta ^{\prime }}\right) ^{2},
\end{eqnarray}%
where $S=L_{x}L_{y}$ and $\beta =\frac{1}{k_{B}T}$ with $k_{B}$ for the
Boltzmann constant. The Fermi-Dirac distribution function is given by $%
f\left( E_{\zeta }\right) =\left[ \exp \left( E_{\zeta }-\mu \right)
/k_{B}T+1\right] ^{-1}$ with the chemical potential $\mu $. The expectation
value of $x_{\zeta }=\left\langle \zeta \left\vert x\right\vert \zeta
\right\rangle $ is evaluated as $x_{\zeta }=$ $\ell _{c}^{2}k_{y}$ and $%
x_{\zeta ^{\prime }}=$ $\ell _{c}^{2}k_{y}^{\prime }$, so that $\left(
x_{\zeta }-x_{\zeta ^{\prime }}\right) ^{2}=\ell _{c}^{4}q_{y}^{2}$ due to $%
k_{y}=k_{y}^{\prime }+q_{y}$ ($q_{y}=q\sin \varphi $ and $%
q^{2}=q_{x}^{2}+q_{y}^{2}$). Since the scattering is elastic and the
eigenvalues do not depend on $k_{y}$, only the transitions $n$ $\rightarrow
n $ are allowed. Conduction occurs by transitions through spatially
separated states from $x_{\zeta }$ to $x_{\zeta ^{\prime }}$, and the
transition rate $W_{\zeta \zeta ^{\prime }}\left( E_{\zeta },E_{\zeta
^{\prime }}\right) $ in the presence of impurities reads%
\begin{eqnarray}
W_{\zeta \zeta ^{\prime }} &=&\sum_{q}\left\vert U_{q}\right\vert
^{2}\left\vert \left\langle \zeta \left\vert e^{i\mathbf{q\cdot r}%
}\right\vert \zeta ^{\prime }\right\rangle \right\vert ^{2}  \notag \\
&&\times \delta \left( E_{\zeta }-E_{\zeta ^{\prime }}\right) .
\end{eqnarray}%
Here, $U_{q}=e^{2}/2\epsilon \sqrt{q^{2}+k_{s}^{2}}$ is the Fourier
transform of the screened impurity potential $U_{r}=\left( e^{2}/4\pi
\epsilon r\right) \exp (-k_{s}r)$, where $k_{s}$ is the screening wave
vector; $\epsilon ={\epsilon }_{0}\epsilon _{r}$ is the dielectric constant.
Now we perform an average over random distribution of impurities and denote $%
N_{I}$ as the impurity density. By virtue of $\Psi _{n,\lambda }^{\tau
_{z}}\left( r\right) \rightarrow \left\vert \zeta \right\rangle $, the $%
W_{\zeta \zeta ^{\prime }}$ is given by%
\begin{eqnarray}
W_{\zeta \zeta ^{\prime }} &=&\frac{2\pi N_{I}}{S\hbar }\sum_{q}\left\vert
U_{q}\right\vert ^{2}\left\vert \varpi _{n,n^{\prime }}(\gamma )\right\vert
^{2}  \notag \\
&&\times \delta \left( E_{\zeta }-E_{\zeta ^{\prime }}\right) \delta
_{k_{y},k_{y}^{\prime }+q_{y}}
\end{eqnarray}%
with%
\begin{equation}
\varpi _{n,n^{\prime }}^{e}(\gamma )=\alpha _{n}\alpha _{n^{\prime
}}J_{n-1,n^{\prime }-1}\left( \gamma \right) +\beta _{n}\beta _{n^{\prime
}}J_{n,n^{\prime }}\left( \gamma \right)
\end{equation}%
for electrons and%
\begin{equation}
\varpi _{n,n^{\prime }}^{h}(\gamma )=\alpha _{n}\alpha _{n^{\prime
}}J_{n,n^{\prime }}\left( \gamma \right) +\beta _{n}\beta _{n^{\prime
}}J_{n-1,n^{\prime }-1}\left( \gamma \right) ,
\end{equation}%
for holes, where $\gamma =$ $l_{c}^{2}q^{2}/2$. The fuctions $\Psi
_{n,\lambda }^{\tau _{z}}\left( r\right) $ oscillating around the point $%
-x_{c}$ allows%
\begin{equation}
\sum_{k_{y}}\rightarrow \frac{L_{y}}{2\pi }\int_{-L_{x}/2\ell
_{c}^{2}}^{L_{x}/2\ell _{\eta }^{2}}dk_{y}=\frac{S}{2\pi \ell _{c}^{2}},
\end{equation}%
and further using cylindrical coordinates%
\begin{equation}
\sum_{q}\rightarrow \frac{S}{4\pi ^{2}l_{c}^{2}}\int_{0}^{2\pi }d\varphi
\int_{0}^{\infty }d\gamma ,
\end{equation}%
we can now evaluate $(5)$ with Eqs. $(6)$, $(10)$ and $(11)$ for elastic
scattering $f(E_{\zeta })=f(E_{\zeta ^{\prime }})$%
\begin{eqnarray}
\sigma _{xx} &=&\frac{N_{I}\beta e^{2}}{\hbar \hbar \omega _{c}}\frac{\left(
e^{2}/2\epsilon \right) ^{2}}{4\pi ^{2}}\sum\limits_{n,\tau _{z}}f(E_{\zeta
})[1-f(E_{\zeta })]  \notag \\
&&\times \int_{0}^{\infty }\int_{0}^{2\pi }\frac{q_{y}^{2}\left\vert \varpi
_{n,n}(\gamma )\right\vert ^{2}}{q^{2}+k_{s}^{2}}d\varphi d\gamma .
\end{eqnarray}%
since $q_{y}=q\sin \varphi $ and $q^{2}=2\gamma /l_{c}^{2}$ as defined
above, Eq.~$(12)$ reads%
\begin{eqnarray}
\sigma _{xx} &=&\frac{N_{I}\beta e^{2}}{\hbar \hbar \omega _{c}}\frac{\left(
e^{2}/2\epsilon \right) ^{2}}{4\pi }\sum\limits_{n,\tau _{z}}f(E_{\zeta
})[1-f(E_{\zeta })]  \notag \\
&&\times \int_{0}^{\infty }\frac{\gamma \left\vert \varpi _{n,n^{\prime
}}(\gamma )\right\vert ^{2}}{\gamma +\gamma _{_{c}}}d\gamma .
\end{eqnarray}%
with $\gamma _{_{c}}=l_{c}^{2}k_{s}^{2}/2$. For the small $q$ limit $q\ll
k_{s}$, $\left( \gamma +\gamma _{_{c}}\right) ^{-1}$ is expanded in powers
of $\gamma /\gamma _{_{c}}$ and we keeps the dominant term%
\begin{eqnarray}
\sigma _{xx} &=&\frac{2N_{I}\beta e^{2}}{\hbar \hbar \omega _{c}}\frac{%
\left( e^{2}/2\epsilon \right) ^{2}}{4\pi l_{c}^{2}k_{s}^{2}}%
\sum\limits_{n,\tau _{z}}f(E_{\zeta })[1-f(E_{\zeta })]  \notag \\
&&\times \int_{0}^{\infty }\gamma \left\vert \varpi _{n,n^{\prime }}(\gamma
)\right\vert ^{2}d\gamma .
\end{eqnarray}%
The calculation of the integral $\int_{0}^{\infty }\gamma \left\vert \varpi
_{n,n^{\prime }}(\gamma )\right\vert ^{2}d\gamma $ in $(14)$ requires%
\begin{equation}
\left\vert J_{n,n^{\prime }}\left( \gamma \right) \right\vert ^{2}=\frac{%
n^{\prime }!}{n!}e^{-\gamma }\gamma ^{n^{\prime }-n}\left[ L_{n^{\prime
}}^{n^{\prime }-n}\left( \gamma \right) \right] ^{2},n\leq n^{\prime }
\end{equation}%
and the functional relations of Laguerre polynomials \cite{29}:%
\begin{equation}
L_{n}^{0}\left( \gamma \right) =L_{n}\left( \gamma \right) ,
\end{equation}%
\begin{equation}
L_{n}^{\alpha -1}\left( \gamma \right) =L_{n}^{\alpha }\left( \gamma \right)
-L_{n-1}^{\alpha }\left( \gamma \right) ,
\end{equation}%
\begin{equation}
xL_{n}^{\alpha +1}\left( \gamma \right) =(n+\alpha +1)L_{n}^{\alpha }\left(
\gamma \right) -(n+1)L_{n+1}^{\alpha }\left( \gamma \right) ,
\end{equation}%
\begin{equation}
\int_{0}^{\infty }e^{-\gamma }\gamma ^{\alpha }L_{n}^{\alpha }\left( \gamma
\right) L_{m}^{\alpha }\left( \gamma \right) d\gamma =\Gamma (\alpha
+n+1)\delta _{nm}/n!.
\end{equation}%
Firstly, making use of $(15)$, one can find some integral identity (e.g.
electrons) for $n=n^{\prime }$:%
\begin{eqnarray}
\int_{0}^{\infty }\gamma \lbrack \beta _{n}^{2}J_{n,n}\left( \gamma \right)
]^{2}d\gamma &=&\int_{0}^{\infty }\gamma \lbrack \beta _{n}^{2}J_{n,n}\left(
\gamma \right) ]^{2}d\gamma \\
&=&\int_{0}^{\infty }\beta _{n}^{4}\gamma e^{-\gamma }\left[ L_{n}\left(
\gamma \right) \right] ^{2}  \notag \\
&=&\int_{0}^{\infty }\beta _{n}^{4}e^{-\gamma }\left[ \gamma L_{n}\left(
\gamma \right) \right] L_{n}\left( \gamma \right) .  \notag
\end{eqnarray}%
Then, with Eqs. $(16)-(18)$, the term of $\gamma L_{n}\left( \gamma \right) $
can be solved
\begin{eqnarray}
\gamma L_{n}\left( \gamma \right) &=&nL_{n}^{-1}\left( \gamma \right)
-(n+1)L_{n+1}^{-1}\left( \gamma \right) \\
&=&n[L_{n}\left( \gamma \right) -L_{n-1}\left( \gamma \right) ]-(n+1)  \notag
\\
&&\times \lbrack L_{n+1}\left( \gamma \right) -L_{n}\left( \gamma \right) ],
\notag
\end{eqnarray}%
Furhther inserting Eqs.~$(21)$ into $(20)$, we finally obtain with Eq.~(19)%
\begin{equation}
\int_{0}^{\infty }\gamma \lbrack \beta _{n}^{2}J_{n,n}\left( \gamma \right)
]^{2}d\gamma =\beta _{n}^{4}\left( 2n+1\right) .
\end{equation}%
In like manner, one can derive%
\begin{equation}
\int_{0}^{\infty }\gamma \lbrack \alpha _{n}^{2}J_{n-1,n-1}\left( \gamma
\right) ]^{2}d\gamma =\alpha _{n}^{4}\left( 2n-1\right)
\end{equation}%
and%
\begin{equation}
\int_{0}^{\infty }2\gamma \alpha _{n}^{2}\beta _{n}^{2}J_{n,n}\left( \gamma
\right) J_{n-1,n-1}\left( \gamma \right) d\gamma =-2n\alpha _{n}^{2}\beta
_{n}^{2}.
\end{equation}%
In a word, since the scattering is elastic and the eigenvalues do not depend
on $k_{y}$, only the transitions $n$ $\rightarrow n$ are allowed, so that%
\begin{eqnarray}
\int_{0}^{\infty }\gamma \left\vert \varpi _{n,n}(\gamma )\right\vert
^{2}d\gamma &=&\left( 2n-1\right) \alpha _{n}^{4}+\left( 2n+1\right)  \notag
\\
&&\times \beta _{n}^{4}-2n\alpha _{n}^{2}\beta _{n}^{2},
\end{eqnarray}%
for $n>0$ and $\left\vert \varpi _{0,0}(\gamma )\right\vert ^{2}=e^{-\gamma
} $ for $n=0$. Finally, we get the results, respectively, for electrons%
\begin{eqnarray}
\sigma _{xx}^{e} &=&\frac{e^{2}}{h}\frac{N_{I}\beta e^{4}}{4\epsilon
^{2}l_{c}^{2}k_{s}^{2}\hbar \omega _{c}}\sum\limits_{n,\tau _{z}}[\left(
2n+1\right) \beta _{n}^{4}+(2n  \notag \\
&&-1)\alpha _{n}^{4}-2n\alpha _{n}^{2}\beta _{n}^{2}]f\left( E_{\zeta
}\right) \left[ 1-f\left( E_{\zeta }\right) \right]
\end{eqnarray}%
and for holes%
\begin{eqnarray}
\sigma _{xx}^{h} &=&\frac{e^{2}}{h}\frac{N_{I}\beta e^{4}}{4\epsilon
^{2}l_{c}^{2}k_{s}^{2}\hbar \omega _{c}}\sum\limits_{n,\tau _{z}}[\left(
2n-1\right) \beta _{n}^{4}+(2n  \notag \\
&&+1)\alpha _{n}^{4}-2n\alpha _{n}^{2}\beta _{n}^{2}]f\left( E_{\zeta
}\right) \left[ 1-f\left( E_{\zeta }\right) \right]
\end{eqnarray}

Figures $1$ exhibits the {collisional} conductivity of the top and bottom
surface states with the SdH periodicity for\emph{\ }zero and\emph{\ }finite
strain. For null strain, the spectra are perfectly symmetric with a single
peak at the CNP confirming the zero-energy TI states (see Fig.~$1$). This
symmetry indicates the surface degeneracy in LLs and thus the SdH
oscillations\ in ${\sigma }_{xx}$ are in phase for different surfaces.
Whereas, with strain applied, the single peak splits two ones with a gap
opening at the CNP and a well resolved beating pattern of SdH oscillations
appears away from the CNP. To clarify these phenomena, we further plot Fig.~$%
2$. As seen, the SdH oscillations\ for different surfaces are out of phase.
This demonstrate that the strain breaks the LLs' surfaces degeneracy as well
with the inversion symmetry of two Dirac cones at both surfaces, which
agrees well with the results of experiments and two Dirac cones model \cite%
{18,19}. Here we remark that such a gap $\left( \Delta _{gap}=2\Delta
\right) $ (see arrow line $1$) does not open between the electrons and holes
for the single Dirac point in each a surface, but between the two Dirac
points at both surfaces. This indicates the two Dirac points at both
surfaces shift in the different directions, suggesting the occurrence of two
inversion asymmetric Dirac cones.

Further analysis reveals that the two peaks in Fig.~$1$ for finite strain
are the superposition of four peaks in Fig.~$2$, i.e., top electron ($\tau
_{+1}^{e})$, top hole ($\tau _{+1}^{h})$, bottom electron ($\tau _{-1}^{e})$%
, and bottom hole ($\tau _{-1}^{h})$. In the same surface, the zero mode
peaks do not split, indicating that the Dirac point at each a surface is not
gapped. However, for the top electrons and bottom holes, the two peaks do
split since the strain lifts the degeneracy of their levels at $n=0$.
Furthermore, for the bottom electrons and top holes, the two peaks not only
split but also exchange (see arrow line $2$ in Fig.~$2$), suggesting a
mixture of LLs. In a word, the strain removes the LLs' degeneracy in
inversion symmetric Dirac cones at both surfaces, and causes the asymmetric
conductivity spectrum presenting different amplitudes or different phases or
both two ones. Such a lifting of degeneracy usually occurs due to Zeeman
coupling in a conventional 2DEG. However, Zeeman coupling here cannot remove
the degeneracy since the inversion symmetry is preserved by the magnetic
field. And the experiments have already definitely excluded the influence of
hybridization between the top and bottom surface states since the width of
surface state ($2-3$ nm) is much smaller than the thickness of sample ($70$%
-nm) \cite{18,19}. So the highly possible mechanism for the degeneracy
lifting (or the LLs splitting) is the inversion asymmetry due to the strain.
Also, from Eq.~$(2)$, one can see the surface degeneracy of LLs is removed
for any nonzero $\Delta $, which makes the mechanism of inversion-symmetry
breaking the most likely explanation for the removed degeneracy. As pointed
out in Ref. \cite{18,19}, the strain induced different electrostatic
environments of both surfaces, breaks the inversion symmetry of two Dirac
cones, lifts the surface degeneracy in LLs, and leads to the asymmetric
conductivity for different surfaces.

The dc Hall conductivity $\sigma _{yx}$ is derived from the nondiagonal
elements of the conductivity tensor as%
\begin{eqnarray}
\sigma _{yx} &=&\frac{i\hbar e^{2}}{S}\sum\limits_{\zeta \neq \zeta ^{\prime
}}f\left( E_{\zeta }\right) \left[ 1-f\left( E_{\zeta ^{\prime }}\right) %
\right] \left\langle \zeta \left\vert \upsilon _{x}\right\vert \zeta
^{\prime }\right\rangle \left\langle \zeta ^{\prime }\left\vert \upsilon
_{y}\right\vert \zeta \right\rangle  \notag \\
&&\times \frac{1-\exp \left( \frac{E_{\zeta }-E_{\zeta ^{\prime }}}{k_{B}T}%
\right) }{\left( E_{\zeta }-E_{\zeta ^{\prime }}\right) ^{2}}.
\end{eqnarray}%
If we use $f\left( E_{\zeta }\right) \left[ 1-f\left( E_{\zeta ^{\prime
}}\right) \right] [1-e^{\beta (E_{\zeta }-E_{\zeta ^{\prime }})}]=f\left(
E_{\zeta }\right) -f\left( E_{\zeta ^{\prime }}\right) $, Eq.~$(31)$ takes
the form of the well known Kubo-Greenwood formula%
\begin{eqnarray}
\sigma _{yx} &=&\frac{i\hbar e^{2}}{S}\sum\limits_{\zeta \neq \zeta ^{\prime
}}\left[ f\left( E_{\zeta }\right) -f\left( E_{\zeta ^{\prime }}\right) %
\right]  \notag \\
&&\times \frac{\left\langle \zeta \left\vert \upsilon _{x}\right\vert \zeta
^{\prime }\right\rangle \left\langle \zeta ^{\prime }\left\vert \upsilon
_{y}\right\vert \zeta \right\rangle }{\left( E_{\zeta }-E_{\zeta ^{\prime
}}\right) ^{2}}.
\end{eqnarray}%
since $\upsilon _{x}=\partial H/\partial p_{x}=-\mathbf{\tau }_{z}\upsilon
_{F}\mathbf{\sigma }_{y}$ and $\upsilon _{y}=\partial H/\partial p_{y}=%
\mathbf{\tau }_{z}\upsilon _{F}\mathbf{\sigma }_{x}$, we get%
\begin{equation}
\left\langle \zeta \left\vert \upsilon _{x}\right\vert \zeta ^{\prime
}\right\rangle =i\mathbf{\tau }_{z}\upsilon _{F}(\alpha _{n}\beta
_{n^{\prime }}\delta _{n-1,n^{\prime }}-\alpha _{n^{\prime }}\beta
_{n}\delta _{n,n^{\prime }-1})
\end{equation}%
and%
\begin{equation}
\left\langle \zeta ^{\prime }\left\vert \upsilon _{y}\right\vert \zeta
\right\rangle =\mathbf{\tau }_{z}\upsilon _{F}(\alpha _{n}\beta _{n^{\prime
}}\delta _{n^{\prime },n-1}+\alpha _{n^{\prime }}\beta _{n}\delta
_{n^{\prime }-1,n}),
\end{equation}%
and thus%
\begin{eqnarray}
I_{nn^{\prime }} &=&\left\langle \zeta \left\vert \upsilon _{x}\right\vert
\zeta ^{\prime }\right\rangle \left\langle \zeta ^{\prime }\left\vert
\upsilon _{y}\right\vert \zeta \right\rangle =i\upsilon _{F}^{2}(\left\vert
\alpha _{n}\beta _{n^{\prime }}\right\vert ^{2}  \notag \\
&&\times \delta _{n-1,n^{\prime }}-\left\vert \alpha _{n^{\prime }}\beta
_{n}\right\vert ^{2}\delta _{n,n^{\prime }-1})
\end{eqnarray}%
As usual the matrix elements between the zeroth level and the other levels
should be treated separately \cite{30}. Corresponding to Eq. $(35)$ one
arrives at%
\begin{equation}
I_{0n^{\prime }}=-i\upsilon _{F}^{2}\left\vert \alpha _{n^{\prime
}}\right\vert ^{2}\delta _{0,n^{\prime }-1},\text{ }I_{n0}=i\upsilon
_{F}^{2}\left\vert \alpha _{n}\right\vert ^{2}\delta _{n-1,0}.
\end{equation}%
In essence, $\sigma _{yx}$ is the diffusive contribution as the collisional
contribution to Hall conductivity vanishes, since the difference of the
matrix elements $x_{\zeta }-x_{\zeta ^{\prime }}=0$. The following
calculations require summing the terms that include all combinations of the
matrix elements $\Sigma _{\lambda \lambda ^{\prime }}$, that is, $\Sigma
_{++}$, $\Sigma _{+-}$, $\Sigma _{-+}$, $\Sigma _{--}$. On the same line
\cite{30}, we gives $\sigma _{\lambda \lambda }=\Sigma _{++}+\Sigma _{--}$ as%
\begin{eqnarray}
\sigma _{yx}^{\lambda \lambda } &=&2A\sum\limits_{n,\tau _{z}}\frac{1}{%
4F_{n}F_{n+1}(E_{n,+}^{\mathbf{\tau }_{z}}-E_{n+1,+}^{\mathbf{\tau }%
_{z}})^{2}}\{[f\left( E_{n,+}^{\mathbf{\tau }_{z}}\right)  \notag \\
&&-f\left( E_{n+1,+}^{\mathbf{\tau }_{z}}\right) +f\left( E_{n,-}^{\mathbf{%
\tau }_{z}}\right) -f\left( E_{n+1,-}^{\mathbf{\tau }_{z}}\right) ]B  \notag
\\
&&+[f\left( E_{n,+}^{\mathbf{\tau }_{z}}\right) -f\left( E_{n+1,+}^{\mathbf{%
\tau }_{z}}\right) +f\left( E_{n+1,-}^{\mathbf{\tau }_{z}}\right)  \notag \\
&&-f\left( E_{n,-}^{\mathbf{\tau }_{z}}\right) ]C\}
\end{eqnarray}%
and $\sigma _{\lambda \lambda ^{\prime }}=\Sigma _{+-}+\Sigma _{-+}$ as%
\begin{eqnarray}
\sigma _{yx}^{\lambda \lambda ^{\prime }} &=&2A\sum\limits_{n,\tau _{z}}%
\frac{1}{4F_{n}F_{n+1}(E_{n,+}^{\mathbf{\tau }_{z}}+E_{n+1,+}^{\mathbf{\tau }%
_{z}})^{2}}\{[f\left( E_{n,+}^{\mathbf{\tau }_{z}}\right)  \notag \\
&&-f\left( E_{n+1,+}^{\mathbf{\tau }_{z}}\right) +f\left( E_{n,-}^{\mathbf{%
\tau }_{z}}\right) -f\left( E_{n+1,-}^{\mathbf{\tau }_{z}}\right) ]D  \notag
\\
&&+[f\left( E_{n,-}^{\mathbf{\tau }_{z}}\right) -f\left( E_{n+1,+}^{\mathbf{%
\tau }_{z}}\right) +f\left( E_{n+1,-}^{\mathbf{\tau }_{z}}\right)  \notag \\
&&-f\left( E_{n,+}^{\mathbf{\tau }_{z}}\right) ]E\}.
\end{eqnarray}%
Here we define $A=\hbar e^{2}\upsilon _{F}^{2}/S$, $B=F_{n}F_{n+1}-\Delta
^{2}$, $C=\tau _{z}\Delta (F_{n}-F_{n+1})$, $D=F_{n}F_{n+1}+\Delta ^{2}$, $%
E=\tau _{z}\Delta (F_{n}+F_{n+1})$. Furthermore, one needs use Eq.~$(10)$,
which indeed gives the degeneracy for each LL and each valley. Notice that
the degeneracy in the zeroth level is only half of that in the nonzero
level, so that we have%
\begin{eqnarray}
\sigma _{yx} &=&\sigma _{yx}^{\lambda \lambda }+\sigma _{yx}^{\lambda
\lambda ^{\prime }}  \notag \\
&=&\sigma _{yx}^{a}+\sigma _{yx}^{b}+\sigma _{yx}^{c}
\end{eqnarray}%
with%
\begin{eqnarray}
\sigma _{yx}^{a} &=&\frac{e^{2}}{h}\sum\limits_{n,\tau _{z}}\left( n+\frac{1%
}{2}\right) [f\left( E_{n,+}^{\mathbf{\tau }_{z}}\right) -  \notag \\
&&f\left( E_{n+1,+}^{\mathbf{\tau }_{z}}\right) +f\left( E_{n,-}^{\mathbf{%
\tau }_{z}}\right) -f\left( E_{n+1,-}^{\mathbf{\tau }_{z}}\right) ],
\end{eqnarray}%
\begin{eqnarray}
\sigma _{yx}^{b} &=&-\frac{e^{2}\Delta ^{2}}{h}\sum\limits_{n,\tau _{z}}%
\frac{\sqrt{n(n+1)}}{F_{n}F_{n+1}}[f\left( E_{n,+}^{\mathbf{\tau }%
_{z}}\right)  \notag \\
&&-f\left( E_{n+1,+}^{\mathbf{\tau }_{z}}\right) +f\left( E_{n,-}^{\mathbf{%
\tau }_{z}}\right) -f\left( E_{n+1,-}^{\mathbf{\tau }_{z}}\right) ],
\end{eqnarray}%
and%
\begin{eqnarray}
\sigma _{yx}^{c} &=&\frac{e^{2}\mathbf{\tau }_{z}\Delta }{8h}%
\sum\limits_{n,\tau _{z}}[\frac{(2n+1)}{F_{n+1}}-\frac{2\sqrt{n(n+1)}}{F_{n}}%
]  \notag \\
&&\times \lbrack f\left( E_{n+1,-}^{\mathbf{\tau }_{z}}\right) -f\left(
E_{n+1,+}^{\mathbf{\tau }_{z}}\right) ]+[\frac{2\sqrt{n(n+1)}}{F_{n+1}}
\notag \\
&&-\frac{(2n+1)}{F_{n}}][f\left( E_{n,+}^{\mathbf{\tau }_{z}}\right)
-f\left( E_{n,-}^{\mathbf{\tau }_{z}}\right) ].
\end{eqnarray}%
In the limit of zero strain and $T\rightarrow 0$, Eq.~$(39)$ can be reduced
to $\sigma _{yx}=\left( 2e^{2}/h\right) (n+\frac{1}{2})$ with the prefactor $%
2$ resulting from the surface degeneracy. Hall plateaus appear at the
filling factor $\pm 1$, $\pm 3$, $\pm 5,$..., agreeing well with the
transport experiments \cite{18,19}.

The collisional and Hall conductivities for different surfaces are both
shown in Fig.~$3$ as a function of the chemical potential for zero and
finite strain. As shown, under strain some extra Hall plateaus arise and the
steps between plateaus coincide with sharp peaks of the collisional
conductivity. Resembling Figs. $1$ and $2$, this originates from the strain
removed surface degeneracy in LLs, so that the density of states could form
the different Landau ladders for different surfaces, thereby causing the
extra quantum plateaus at even filling factors $0$, $\pm 2$, $\pm 4$, $\pm 6$%
, $\pm 8$, ..., etc.\emph{\ }As well known, the Hall conductivity of a
single Dirac cone in strainless graphene is given by $\sigma _{yx}=\left(
2n+1\right) e^{2}/h$ with the odd filling factors $2n+1$, directly leading
to the total filling factors $\left( 4n+2\right) $ (Ref. \cite{31}). It
seems that the new Hall plateaus for a single Dirac cone arise from the
extra value $\sigma _{xy}=ve^{2}/h$ with the filling factors $0$, $\pm 1$, $%
\pm 2$, $\pm 3$, ..., etc. In fact, these values are nonexistent, and the
abnormal Hall behaviour should be attributed to the phase difference of Hall
conductivity for different surfaces induced by the strain.

\section{Conclusions}

In conclusion, we have investigated the strain effects on the quantum
magnetotransport properties for the surface states of TIs at finite
temperature and magnetic field. The strain are shown to remove the surface
degeneracy of LLs in the two inversion symmetric Dirac cones. Thus, the
Dirac particles of different surfaces present the well seperated quantum
Hall and SdH effects with different amplitudes and phases. This accordingly
gives rise to the extra Hall plateaus and the SdH beating pattern away from
the CNP. In addition, the SdH conductivity under strain possess two
zero-mode peaks around the CNP, while for null strain there is just a single
CNP peak. We interpret the two peaks are the superposition of four peaks
arising from top electron, top hole, bottom electron, and bottom hole.
Further analysis reveals that, in the same surface, the zero mode peaks do
not split, indicating the Dirac point at each a surface is not gapped.
However, for the top electrons and bottom holes, the two peaks do split
since the strain lifts the degeneracy of their levels at $n=0$. Furthermore,
for the bottom electron and top hole, the two peaks not only split but also
exchange, suggesting a mixture of LLs. These should be sufficient to well
appreciate the experimental results on the quantum magnetotransport of the
surface states of HgTe.

\section{ACKNOWLEDGMENTS}

This work is financially supported by the National Nature Science Foundation
of China (Grant Nos. 11074196 and 11304241). We also acknowledges support
from the School Foundation (Grant No. 1205-04020102) and the Qualified
Personnel Foundation of Taiyuan University of Technology(QPFT) (Nos.
tyutrc-201273a).\newline

\end{document}